\documentclass[11pt,eqs]{article}
\usepackage{latexsym}
\usepackage{amsmath}
\usepackage{hyperref}
\hypersetup{
	colorlinks=true,
	linkcolor=cyan,
	filecolor=magenta,      
	urlcolor=cyan,
}

\makeatletter
\newcommand\xleftrightarrow[2][]{%
	\ext@arrow 9999{\longleftrightarrowfill@}{#1}{#2}}
\newcommand\longleftrightarrowfill@{%
	\arrowfill@\leftarrow\relbar\rightarrow}
\makeatother

\title{
	Noncommutativity and nonassociativity of type II superstring with coordinate dependent RR field - the general case
	\thanks{Work supported in part by the Serbian Ministry of Education,Science and Technological Development.}}
\author{D. Obri\'c and B. Nikoli\'c \thanks{email:dobric, bnikolic@ipb.ac.rs}\\{\it Institute of Physics Belgrade, University of Belgrade, Pregrevica 118, Serbia} }

\begin{document}
	
	\maketitle
	
\begin{abstract}

In this paper we consider non-commutativity that arises from T-duality of bosonic coordinates of type II superstring in presence of coordinate dependent Ramond-Ramond field.  Action with such choice of the background fields is not translational invariant. Consequently, we will employ generalization of Buscher procedure that can be applied to cases that have coordinate dependent fields and that do not possess translational isometry. Bosonic part of newly obtained T-dual theory is non-local and defined in non-geometric double space spanned by Lagrange multipliers $y_\mu$ and double coordinate $\Delta V^\mu$. We will apply Buscher procedure once more on T-dual theory to check if original theory can  be salvaged. Finally, we will use T-dual transformation laws along with Poisson brackets of original theory to derive Poisson bracket structure of T-dual theory.		
		
\end{abstract}

\section{Introduction}
\setcounter{equation}{0}

String theory as a possible candidate for unification of all known interactions offers a framework for description of both gauge interactions and gravity. Analyzing the relation between world-sheet diffeomorphisms and transformations of the background fields for open bosonic string \cite{mebo1, mebo2, ljdbs} it is concluded that Kalb-Ramond field gets one additional term that is in fact a field strength of some gauge field. In sigma-model action it looks like that gauge fields are attached at the string endpoints moving along $Dp$-brane. Finiteness of the gauge theories (UV cutoff) demands existence of some minimal length. Consequently, noncommutativity naturally arose in open bosonic string theory in the presence of the constant background fields \cite{open string non-commutativity, sazda1, sazda2, sazda3}.

The fact that noncommutativity appears together with gauge theory produces a new line of investigation in quantum field theory - noncommutative gauge theories \cite{nc1, nc2, nc3}, but not noncommutative gravity. 

The open bosonic string in the presence of the constant background fields gives constant noncommutativity \cite{open string non-commutativity, sazda1, sazda2, sazda3}, but, consequently, Jacobi identity is zero and associativity is not broken. Closed string in the presence of the constant background fields remains commutative.

Noncommutativity in open string theory comes from the boundary conditions (see \cite{sazda1, sazda2, sazda3}). Coordinates and their canonically conjugated momenta are mixed in the boundary conditions, and because they obey standard Poisson algebra, at the end we get noncommutativity of the initial coordinates. All these facts tell us about the way how we can reach noncommutativity in the bosonic closed string case. Kalb-Ramond field must be, at least, linearly coordinate dependent, and the generalized T-dualization procedure is a machinery \cite{T-duality explained 1, T-duality explained 2, Luest T-duality and geometry, T-duality procedure 1, T-duality procedure 2, Generalized Buscher procedure, Gereralized Buscher procedure 2, Generalized Buscher procedure 3, Auxiliary action Buscher procedure}. The noncommutativity relations are coordinate dependent and produce nonzero Jacobi identity - nonassociativity appears in the closed string theory \cite{Luest T-duality and geometry, Inversion of equations of motion, Generalized Buscher procedure 3, nasrad1, nasrad2}.

The noncommutativity and nonassociativity can be considered also within superstring theory \cite{book1, book2}. In Ref.\cite{nasrad3} we considered one special case of the type II superstring theory in pure spinor formulation \cite{pure spinor formalism papers 1, pure spinor formalism papers 2, pure spinor formalism papers 3, pure spinor formalism papers 4, Vertex operators} - all physical background fields are constant except Ramond-Ramond (RR) field strength. The RR field strength consists of the constant part and linearly coordinate dependent one, which is infinitesimal. In accordance with consistency conditions, we have chosen constant part of RR field strength to be symmetric and cooordinate dependent part to be antisymmetric tensor. 

The motivation for this choice of background fields is the quest for the anticommutation relation between fermionic coordinates suggested in \cite{Vertex operators, original paper 2}. Formally, this case is similar to the bosonic string case with coordinate dependent Kalb-Ramond field (weakly curved background). The difference is that in the superstring case noncommutativity parameter depends both on the bosonic and fermionic coordinates. Also we obtained that Jacobiator is nonzero. Both noncommutativity and nonassociativity parameters are proportional to the infinitesimal tensor from RR field strength.

In this article we consider the same action as in \cite{nasrad3}, but we will not imply the additional restrictions on the constant and coordinate dependent part of RR field strength as in \cite{nasrad3}. The fundamental difference in relation to the choice of background field in \cite{nasrad3} is in the fact that action with RR field strength without restrictions does not possess translational isometry. In that sense this case can be considered as general one comparing with \cite{nasrad3}.

We will use the generalized T-dualization procedure \cite{Generalized Buscher procedure, Auxiliary action Buscher procedure} along bosonic directions. Because this general case cannot be deduced to the form of the bosonic string with linearly dependent Kalb-Ramond field, as it could in the case \cite{nasrad3}, we obtained more complicated form of T-dual transformation laws and, consequently, the generalization of $\beta_\mu$ functions in the form of $N(\xi)$ functions. Besides the complexity of the T-dual transformation laws we succeeded to find expressions for noncommutativity and nonassociativity as well as the form of the T-dual theory.

At the end we give some concluding remarks.
In the appendices we present the derivation of $N(\xi)$ functions and show their properties.

\section{General type II superstring action and  choice of background fields}
\setcounter{equation}{0}

In this section we will shortly present how we derive the action of type II superstring in pure spinor formulation with all constant background fields except RR field strength from the general form of that action given in \cite{Vertex operators, pure spinor formalism papers 1, pure spinor formalism papers 2, pure spinor formalism papers 3, pure spinor formalism papers 4}. 

\subsection{General form of the pure spinor type II superstring action}

The general form of the type II superstring action in pure spinor formalism is derived and given in \cite{Vertex operators}. It consists of two parts and can be represent as their sum
\begin{equation}
S=S_0 + V_{SG}\, ,
\end{equation}
where $S_0$ describes the motion of string in flat background 
\begin{equation} \label{S_0 action}
S_0 = \int_{\Sigma}d^2 \xi \left( \frac{k}{2} \eta_{ \mu \nu } \partial_m x^\mu \partial_n x^\nu \eta^{m n} - \pi_\alpha \partial_- \theta^\alpha + \partial_+ \bar{\theta}^\alpha \bar{\pi}_\alpha \right) 
+S_\lambda +S_{ \bar{\lambda} }\, ,
\end{equation}
while the second one contains all possible interactions
\begin{equation}
V_{SG} =  \int_{\Sigma}d^2 \xi (X^T)^M A_{MN} \bar{X}^N.
\end{equation}
The second part of the action is expressed in terms of the integrated form of massless type II supergravity vertex operator $V_{SG}$. The actions $S_{\lambda}$ and $S_{\bar{\lambda}}$ in (\ref{S_0 action}) are {free-field} actions for pure spinors
 \begin{equation}
 S_\lambda = \int d^2 \xi \omega_\alpha \partial_- \lambda^\alpha, \quad S_{\bar{\lambda}} = \int d^2 \xi \bar{\omega}_\alpha \partial_+ \bar{\lambda}^\alpha\, .
 \end{equation}
Here, $\lambda^\alpha$ and $\bar{\lambda}^\alpha$ are pure spinors whose canonically conjugated momenta are $\omega_\alpha$ and $\bar{\omega}_\alpha$, respectively.

The vectors $X^M$ and $X^N$ and matrix $A_{MN}$ are of the form
\begin{equation}
X^M = \left(\begin{matrix}
\partial_+ \theta^\alpha\\
\Pi_+^\mu\\
d_\alpha\\
\frac{1}{2} N_+^{\mu \nu}
\end{matrix} \right), 
\quad \bar{X}^M = \left( \begin{matrix}
\partial_- \bar{\theta}^\lambda\\
\Pi_-^\mu\\
\bar{d}_\lambda\\
\frac{1}{2} \bar{N}_-^{\mu \nu}
\end{matrix} \right), \quad
A_{MN} = \begin{bmatrix}
A_{\alpha \beta} & A_{\alpha \nu} & {E_\alpha}^\beta & \Omega_{\alpha, \mu \nu} \\
A_{\mu \beta} & A_{\mu \nu} & \bar{E}_\mu^\beta & \Omega_{\mu, \nu \rho}\\
{E^\alpha}_\beta & E_\nu^\alpha & P^{\alpha \beta} & {C^\alpha}{}_{\mu \nu}\\
\Omega_{\mu \nu, \beta} & \Omega_{\mu \nu, \rho} & {\bar{C}^{\beta}{}_{\mu \nu}} & S_{\mu \nu, \rho \sigma}
\end{bmatrix},
\end{equation} 
where notation is taken from Refs.\cite{Vertex operators, nasrad3}. Every component of the matrix $A_{MN}$ is function of bosonic, $x^\mu$, and fermionic, $\theta^\alpha$ and $\bar{\theta}^\alpha$, coordinates. For more details about derivation of the components consult \cite{Vertex operators}. The superfields $A_{\mu \nu}$, $\bar{E}_\mu^\alpha$, $E_\mu^\alpha$ and $P^{\alpha \beta}$ are known as physical superfields, superfields that are in the first row and the first column are known as auxiliary because they can be expressed in terms of physical ones \cite{Vertex operators}. Remaining superfields $\Omega_{\mu, \nu \rho}\  (\Omega_{\mu \nu, \rho})$, $C^\alpha{}_{\mu \nu}\  ({\bar{C}^\beta{}_{\mu\nu}})$ and $S_{\mu\nu,\rho\sigma}$, are curvatures (field strengths) for physical fields.   Components of $X^M$ and $\bar{X}^N$ are of the form
\begin{equation}
\Pi_+^\mu = \partial_+ x^\mu + \frac{1}{2}\theta^\alpha (\Gamma^\mu)_{\alpha \beta}\partial_+\theta^\beta, \quad  
\Pi_-^\mu = \partial_- x^\mu + \frac{1}{2} \bar{\theta}^\alpha (\Gamma^\mu)_{\alpha \beta} \partial_- \bar{\theta}^\beta,
\end{equation}

\begin{align}
d_\alpha &= \pi_\alpha - \frac{1}{2} (\Gamma_\mu \theta)_\alpha \left[ \partial_+ x^\mu +\frac{1}{4} (\theta \Gamma^\mu \partial_+ \theta)
\right],  \nonumber\\
\bar{d}_\alpha &= \bar{\pi}_\alpha - \frac{1}{2}(\Gamma_\mu \bar{\theta})_\alpha \left[ \partial_- x^\mu +\frac{1}{4}(\bar{\theta} \Gamma^\mu \partial_- \bar{\theta} )
\right],
\end{align}

\begin{equation}
N_+^{\mu \nu} = \frac{1}{2} \omega_\alpha {( \Gamma^{[\mu\nu]}
)^\alpha}_\beta \lambda ^\beta, \quad \bar{N}_-^{\mu\nu} = \frac{1}{2} \bar{\omega}_\alpha {(  \Gamma^{ [\mu\nu ]  }     )^\alpha}_\beta \bar{\lambda}^\beta.
\end{equation}

 The world-sheet is spanned by $\xi^m = (\xi^0 = \tau, \xi^1 = \sigma)$, while world-sheet light-cone partial derivatives are defined as $\partial_\pm = \partial_\tau \pm \partial_\sigma$. Superspace contains bosonic $x^\mu \  (\mu = 0,1,...,9)$ and fermionic $\theta^\alpha,\  \bar{\theta}^\alpha \  (\alpha = 1,2,...,16)$ coordinates. Variables $\pi_\alpha$ and $\bar{\pi}_\alpha$ are canonically conjugated momenta to the fermionic coordinates $\theta^\alpha$ and $\bar{\theta}^\alpha$, respectively.

 \subsection{Choice of the background fields}
 
 In this particular case we will use the supermatrix $A_{MN}$ where all physical background fields, except RR field strength $P^{\alpha \beta}$, are constant. RR fields strength will have linear coordinate dependence on bosonic coordinate $x^\mu$. Consequently, supermatrix $A_{MN}$ is of the following form
 
 \begin{equation}
 A_{MN} = \begin{bmatrix}
 0 & 0 & 0 & 0 \\
 0 & k(\frac{1}{2}g_{\mu \nu} + B_{\mu \nu}) & \bar{\Psi}_\mu^\beta & 0\\
 0 & -\Psi_\nu^\alpha & \frac{2}{k}(f^{\alpha\beta} + C_\rho^{\alpha\beta} x^\rho)& 0\\
 0 & 0 & 0 & 0
 \end{bmatrix},
 \end{equation}  
where $g_{\mu \nu}$ is symmetric tensor, $B_{\mu \nu}$ is Kalb-Ramond antisymmetric field, $\Psi_\mu^\alpha$ and $\bar{\Psi}_\mu^\alpha$ are Mayorana-Weyl gravitino fields and $f^{\alpha \beta}$ and $C_\rho^{\alpha \beta}$ are constant tensors. Let us stress this will be a classical analysis and we will not calculate the dilaton shift under T-duality transformation.
 
From the consistency conditions given in Ref.\cite{Vertex operators}, following this choice of background fields, it follows
\begin{align} \label{constraint}
\gamma^\mu_{\alpha \beta} C_\mu^{\beta \gamma} = 0, \quad \gamma^\mu_{\alpha \beta} C_\mu^{\gamma \beta} = 0.
 \end{align}
 
Because all background fields are expanded in powers of $\theta^\alpha$ and $\bar\theta^\alpha$, $\theta^{\alpha}$ and $\bar{\theta}^{\alpha}$ terms in $X^M$ and $\bar{X}^N$ will be neglected. Taking into account all imposed assumptions and approximations, the full action $S$ is getting the form
\begin{equation}\label{Action full S}
\begin{gathered} 
S        =          \int_{\Sigma}   d^2   \xi          \left[     \frac{k}{2}   \Pi_{ +\mu \nu } \partial_+   x^\mu  \partial_-   x^\nu          -           \pi_\alpha  ( \partial_-   \theta^\alpha   +   \Psi_\nu^\alpha   \partial_-  x^\nu )   +              (   \partial_+   \bar{ \theta }^\alpha +   \partial_+   x^\mu \bar{ \Psi }_\mu^\alpha  )   \bar{ \pi }_\alpha    \right.\\
\left.             +          \frac{2}{k} \pi_\alpha   (   f^{ \alpha\beta }   +C_\rho^{\alpha\beta} x^\rho   ) \bar{ \pi }_\beta          \right],
\end{gathered}
\end{equation}
where $\Pi_{ \pm \mu \nu } = B_{\mu \nu} \pm \frac{1}{2} G_{\mu \nu}$, and $G_{\mu \nu} = \eta_{\mu \nu} + g_{\mu \nu}$ is metric tensor. The actions $S_{\lambda}$ and $S_{\bar{\lambda}}$ are fully decoupled from the rest and they will not be analyzed from now on.

It is easy to notice that fermionic momenta play the roles of the auxiliary fields in full action. They can be integrated out finding equations for motion for both $\pi_\alpha$ and $\bar{ \pi }_\alpha$

\begin{alignat}{2} \label{Equations for fermionic momenta 1}
\bar{ \pi }_\beta   &   =   
&&\frac{k}{2}   
\left( F^{-1}  (x)   \right)_{ \beta \alpha}        
\left(       \partial_- \theta^\alpha     +    \Psi_\nu^\alpha    \partial_-   x^\nu     \right) , \\
\label{Equations for fermionic momenta 2}
\pi_\alpha    &   =    
-   &&\frac{k}{2}
\left(     \partial_+   \bar{ \theta }^\beta     +     \partial_+   x^\mu    \bar{ \Psi }_\mu^\beta   \right)
\left( F^{-1}  (x)   \right)_{ \beta \alpha} \, ,
\end{alignat}
where $F^{\alpha\beta}(x)$ and $(F^{-1}(x))_{\alpha\beta}$ are of the form 
\begin{equation}
F^{\alpha\beta}(x)=f^{\alpha\beta}+C_\mu^{\alpha\beta} x^\mu\, ,\quad (F^{-1}(x))_{\alpha\beta}=( f^{ -1 } )_{ \alpha\beta }  - ( f^{ -1 } )_{ \alpha\alpha_1 }   C_\rho^{\alpha_1 \beta_1}    x^\rho   ( f^{ -1 } )_{ \beta_1\beta }\, .
\end{equation}
For practical reasons, we assume that $C^{\alpha\beta}{}_\mu$ is infinitesimal. This assumption is in accordance with constraints (\ref{constraint}).
Substituting equations (\ref{Equations for fermionic momenta 1}) and (\ref{Equations for fermionic momenta 2}) into (\ref{Action full S}) the final form of action is

\begin{align} \label{Action final S}
S       =    
k    \int_{\Sigma}    d^2    \xi      
\left[     \Pi_{ +\mu \nu }   \partial_+     x^\mu    \partial_-   x^\nu
+     \frac{1}{2}      (  \partial_+  \bar{ \theta }^\alpha  +   \partial_+   x^\mu   \bar{ \Psi }_\mu^\alpha     )
\left( F^{-1}  (x)   \right)_{ \alpha\beta }
(    \partial_-    \theta^\beta      + \Psi_\nu^\beta      \partial_-  x^\nu         )
\right] .  
\end{align}
Let us note that we did not impose any conditions on tensors $f^{\alpha\beta}$ and $C^{\alpha\beta}_\mu$ as we did in Ref.\cite{nasrad3}. The case considered in this article is more general than the case studied in \cite{nasrad3}, because action does not possess translational symmetry.

\section{T-dualization}
\setcounter{equation}{0}

Here we will make T-dualization of all bosonic directions aiming to find T-dual transformation laws - relations between T-dual coordinates and canonical variables of the original theory. The T-dual transformation laws will be used to calculate Poisson brackets of the T-dual coordinates. 

\subsection{Implementation of the generalized T-dualization procedure}

In implementing of T-dualization procedure we will use {\it generalized Buscher T-dualization procedure} \cite{Generalized Buscher procedure}. The standard Buscher procedure \cite{T-duality procedure 1,T-duality procedure 2} is made to be used along directions on which background fields do not depend (isometry directions), while generalized Buscher procedure can be applied to theories with coordinate dependent background fields along all directions. The generalized T-dualization procedure follows three steps - introduction of covariant derivatives, invariant coordinates and additional gauge fields, which produces additional degrees of freedom. The starting and T-dual theory must have the same number of degrees of freedom. In order to achieve that we eliminate all excessive degrees of freedom demanding that field strength of gauge fields $(F_{+ -} = \partial_+ v_- - \partial_- v_+)$ vanishes by addition of Lagrange multipliers. Then we fix the gauge symmetry (shift symmetry) and action is left with gauge fields and their derivatives. Finding equations of motion for gauge fields, expressing in terms of the Lagrange multipliers and inserting those equations into action we obtain T-dual action, where Lagrange multipliers have roles of T-dual coordinates.  

T-duality can be performed also in the cases of the absence of shift symmetry \cite{Auxiliary action Buscher procedure}. Then we replace original action with translation invariant auxiliary action. Form of the auxiliary action is exactly the same as the form of action where translation symmetry was localized and gauge fixed. It produces correct T-dual theory only if original action can be salvaged from it.

Action (\ref{Action final S}) is not translational invariant. Consequently, we make the following substitutions

\begin{alignat}{2}
&\partial_\pm x^\mu     &&\rightarrow       v_\pm^\mu, \\ 
&x^\rho                  &&\rightarrow       \Delta V^\rho    =     \int_{P} d { \xi^\prime }^m v_m^\rho (\xi^\prime)\label{V expresion},\\ 
&S                    &&\rightarrow     S  +  \frac{k}{2}    \int_{\Sigma}       d^2       \xi          \left[ v_+^\mu \partial_- y_\mu   -  v_-^\mu   \partial_+ y_\mu
\right],
\end{alignat}
and insert them in action (\ref{Action final S}). The result is auxiliary action convenient for T-dualization procedure
\begin{equation}\label{Auxiliary action S_{aux}}
\begin{gathered}  
S_{aux}       =    k    \int_{\Sigma}    d^2    \xi
\left[ \Pi_{ +\mu \nu } v_+^\mu v_-^\nu    +    \frac{1}{2} (      \partial_+     \bar{ \theta }^\alpha      +     v_+^\mu    \bar{ \Psi }_\mu^\alpha     )         \left( F^{-1}  ( \Delta V )   \right)_{ \alpha\beta }   (      \partial_-\theta^\beta     + \Psi_\nu^\beta  v_-^\nu       ) \right. \\
\left.  +      \frac{1}{2}(     v_+^\mu    \partial_- y_\mu           -         v_-^\mu       \partial_+      y_\mu       )
\right].
\end{gathered}
\end{equation}
Let us note that path $P$ starts from $\xi_0$ and ends in $\xi$. In this way action becomes non-local.

Finding equations of motion for Lagrange multipliers

\begin{equation} \label{x form v}
\partial_- v_+^\mu - \partial_+ v_-^\mu = 0 \quad v_\pm^\mu = \partial_\pm x^\mu,
\end{equation}
and inserting them into (\ref{V expresion}) we have
\begin{equation} \label{X from V}
\Delta V^\rho = \int_{P} d { \xi^\prime }^m \partial_m x^\rho (\xi^\prime) = x^\rho (\xi) - x^\rho (\xi_0) = \Delta x^\rho.
\end{equation}
In absence of translational symmetry, in order to extract starting action from auxiliary one, we impose $x^\rho (\xi_0) =0$ as a constraint. Taking all this into account, we get the starting action (\ref{Action final S}).

Euler-Lagrange equations of motion for gauge fields $v_\pm(\kappa)$ give the following ones

\begin{align} \label{eq of motion for v_+}
\begin{gathered} 
-\frac{1}{2}   \partial_- y_\mu  (\kappa)   =   
\Pi_{ +\mu \nu }   v_-^\nu (\kappa)  +
\frac{1}{2}  \bar{ \Psi }_\mu^\alpha     \left( F^{-1}  ( \Delta V )   \right)_{ \alpha\beta }
(    \partial_-  \theta^\beta  (\kappa)     +  \Psi_\nu^\beta v_-^\nu  (\kappa)  )  \\
- \frac{1}{2}        \int_{\Sigma}           d^2          \xi
\big[    \partial_+     \bar{ \theta }^\alpha (\xi)     +     v_+^{\nu_1}  (\xi)   \bar{ \Psi }_{\nu_1} ^\alpha    \big]  
( f^{ -1 } )_{ \alpha\alpha_1 }     C_\mu^{\alpha_1 \beta_1}      ( f^{ -1 } )_{ \beta_1\beta }
N (\kappa^+) \big[     \partial_-\theta^\beta (\xi)    + \Psi_{\nu_2} ^\beta  v_-^{\nu_2}   (\xi)     \big], \\
\end{gathered}\\
\begin{gathered}
\frac{1}{2} \partial_+ y_\mu  (\kappa)  =
\Pi_{ +\nu \mu }  v_+^\nu (\kappa)
+\frac{1}{2}    (    \partial_+  \bar{ \theta }^\alpha  (\kappa)  +  v_+^\nu (\kappa)  \bar{ \Psi }_\nu^\alpha  )    \left( F^{-1}  ( \Delta V )   \right)_{ \alpha\beta }   \Psi_\mu^\beta  \\
- \frac{1}{2}        \int_{\Sigma}           d^2          \xi
\big[    \partial_+     \bar{ \theta }^\alpha (\xi)     +     v_+^{\nu_1}  (\xi)   \bar{ \Psi }_{\nu_1} ^\alpha    \big]  
( f^{ -1 } )_{ \alpha\alpha_1 }     C_\mu^{\alpha_1 \beta_1}      ( f^{ -1 } )_{ \beta_1\beta }
N (\kappa^-)  \big[     \partial_-\theta^\beta (\xi)    + \Psi_{\nu_2} ^\beta  v_-^{\nu_2}  (\xi)     \big]. 
\end{gathered} \label{eq of motion for v_-}
\end{align}
Here, function $N(\kappa^\pm)$ is obtained from variation of term containing $\Delta V^\rho$ in expression for $ F^{-1}  ( \Delta V )$ (details are presented in Appendix \ref{appendix A}). They represent the generalization of beta functions introduced in Ref.\cite{nasrad3}

\begin{align}
N({\kappa^+}) =& 
\delta      \Big(      { \xi^\prime }^- \big( ( { { \xi^\prime }^+ })^{-1}
(\kappa^+) \big)   - \kappa^-
\Big)     
\left[    H( \xi^+   - \kappa^+  )  -  H(\xi_0^+    -   \kappa^+)    \right], \\
N(\kappa^-) 
=&  
\delta      \Big(      { \xi^\prime }^+ \big( ( { { \xi^\prime }^- })^{-1}
(\kappa^-) \big)   - \kappa^+
\Big)     
\left[    H( \xi^-   - \kappa^-  )  -  H(\xi_0^-    -   \kappa^-)    \right],
\end{align}
where more details on Dirac delta function and step function are given in Appendix \ref{appendix A}. As we see the expressions for derivatives of $y_\mu$ are more complex comparing with those in \cite{nasrad3}, where translational symmetry is present.

Assuming that $C_\mu^{\alpha \beta}$ is an infinitesimal, we can iteratively invert equations of motion (\ref{eq of motion for v_+}) and (\ref{eq of motion for v_-}) \cite{Inversion of equations of motion}. Separating variables into two parts, one finite and one infinitesimal proportional to $C_\mu^{\alpha \beta}$, we have



\begin{align}
\label{v_-^n as function of y}
\begin{gathered}
v_-^\nu  (\kappa)        = 
-   \frac{1}{2}  \bar{\Theta}_-^{\nu \nu_1} 
\Bigg\{ \Bigg.
\partial_-     y_{\nu_1}  (\kappa)       +     \bar{ \Psi }_{\nu_1}^\alpha     ( F^{ -1 } ( \Delta V  ) )_{\alpha\beta}   \partial_-   \theta^\beta (\kappa)     
 \\
\begin{aligned}
&+ \frac{1}{2}  \Psi_{\nu_1}^\alpha  ( f^{- 1 } )_{ \alpha\alpha_1 }  C_\rho^{\alpha_1 \alpha_2} \Delta V^\rho ( f^{- 1 } )_{ \alpha_2\alpha_3 }     \Psi_{\nu_2}^{\alpha_3}   \bar{\Theta}_-^{\nu_2 \nu_3} 
\Big(   \partial_-  y_{\nu_3} (\kappa)   + \bar{ \Psi }_{\nu_3} ^{ \beta_1 }     ( f^{- 1 } )_{ \beta_1\beta }      \partial_-    \theta^{ \beta } (\kappa)  \Big)   \\
&-       \int_{\Sigma}          d^2 \xi
\Big[    \partial_+     \bar{ \theta }^\alpha (\xi)     +     \frac{1}{2} \Big( \partial_+ y_{\mu_1} (\xi) -  \partial_+ \bar{ \theta }^{\gamma_1} (\xi)  ( f^{- 1 } )_{ \gamma_1\gamma_2 } \Psi_{\mu_1}^{\gamma_2} 
\Big)  \bar{\Theta}_-^{\mu_1 \nu_1}
\bar{ \Psi }_{\nu_1} ^\alpha    \Big] ( f^{ -1 } )_{ \alpha\alpha_1 }     C_{\nu_1}^{\alpha_1 \beta_1}  \\
&\quad\times ( f^{ -1 } )_{ \beta_1\beta }
N (\kappa^+)
\Big[     \partial_-\theta^\beta (\xi)    -\frac{1}{2} \Psi_{\nu_2} ^\beta \bar{\Theta}_-^{\nu_2 \mu_2} 
\Big( \partial_- y_{\mu_2} (\xi) + \bar{ \Psi }_{\mu_2}^{\gamma_3}  ( f^{- 1 } )_{ \gamma_3\gamma_4 } \partial_- \theta^{\gamma_4}  (\xi)
\Big)
\Big] \Bigg. \Bigg\},
\end{aligned}
\end{gathered} \\
\label{v_+^n as function of y}
\begin{gathered}
 v_+^\mu (\kappa)  = 
\frac{1}{2} \bar{\Theta}_-^{\mu_1 \mu}   
\Bigg\{ \Bigg.
\partial_+   y_{\mu_1}    (\kappa)   -      \partial_+ \bar{ \theta }^\alpha (\kappa) ( F^{ -1 } ( \Delta V  ) )_{\alpha\beta} \Psi_{\mu_1}^\beta 
 \\
\begin{aligned}
&+\frac{1}{2} \Big( \partial_+    y_{\mu_2} (\kappa)     -   \partial_+   \bar{ \theta }^\alpha (\kappa)  ( f^{- 1 } )_{ \alpha\alpha_1 } \Psi_{\mu_2}^{\alpha_1}     \Big)
\bar{\Theta}_-^{\mu_2 \mu_3} 
\bar{ \Psi }_{\mu_3}^{\beta_3}     ( f^{- 1 } )_{ \beta_3\beta_2 } C_\rho^{\beta_2 \beta_1} \Delta V^\rho ( f^{- 1 } )_{ \beta_1\beta }     \Psi_{\mu_1}^{\beta}  \\
&+       \int_{\Sigma}          d^2 \xi
\Big[    \partial_+     \bar{ \theta }^\alpha (\xi)     +     \frac{1}{2} \Big( \partial_+ y_{\mu_2} (\xi) -  \partial_+ \bar{ \theta }^{\gamma_1} (\xi)  ( f^{- 1 } )_{ \gamma_1\gamma_2 } \Psi_{\mu_2}^{\gamma_2} 
\Big)  \bar{\Theta}_-^{\mu_2 \nu_1}
\bar{ \Psi }_{\nu_1} ^\alpha    \Big] ( f^{ -1 } )_{ \alpha\alpha_1 }     C_{\mu_1}^{\alpha_1 \beta_1}     \\
&\quad\times ( f^{ -1 } )_{ \beta_1\beta }
N (\kappa^-)
\Big[     \partial_-\theta^\beta (\xi)    -\frac{1}{2} \Psi_{\nu_2} ^\beta \bar{\Theta}_-^{\nu_2 \mu_3} 
\Big( \partial_- y_{\mu_3} (\xi) + \bar{ \Psi }_{\mu_3}^{\gamma_3}  ( f^{- 1 } )_{ \gamma_3\gamma_4 } \partial_- \theta^{\gamma_4}  (\xi)
\Big)
\Big] \Bigg. \Bigg\}.
\end{aligned}
\end{gathered}
\end{align}
Tensor $\bar{\Theta}_-^{\mu\nu}$ is inverse tensor to $\bar{\Pi}_{+\mu\nu} = \Pi_{ +\mu \nu } + \frac{1}{2} \bar{ \Psi }_\mu^\alpha (f^{-1})_{\alpha\beta}\Psi_\nu^\beta $

\begin{equation}
\bar{\Theta}_-^{\mu\nu} \bar{\Pi}_{+\nu\rho} = \delta^\mu_\rho,
\end{equation}
where
\begin{gather}
\bar{\Theta}_-^{\mu\nu} = \Theta_-^{\mu\nu} - \frac{1}{2} \Theta_-^{\mu\mu_1} \bar{ \Psi }_{\mu_1}^\alpha (\bar{f}^{-1})_{\alpha\beta} \Psi^\beta_{\nu_1} \Theta_-^{\nu_1\nu},\\
\bar{f}^{\alpha\beta} = f^{\alpha \beta} + \frac{1}{2} \Psi_\mu^\alpha \Theta_-^{\mu\nu} \bar{ \Psi }_\nu^\beta,\\
\Theta_-^{\mu\nu} \Pi_{ +\mu \rho } = \delta^\mu_\rho, \qquad \Theta_- = -4 (G_E^{-1} \Pi_- G^{-1})^{\mu\nu}.
\end{gather}
Effective metric tensor is defined as $G_{E \mu\nu} \equiv G_{\mu \nu} -4(BG^{-1}B)_{\mu\nu} $.

In above expressions $\Delta V$ is a quantity in the zeroth order in $C^{\alpha\beta}_\mu$

\begin{align}
\begin{gathered}
\Delta V^\rho  =  \int d\xi^+ v_+^\rho+\int d\xi^- v_-^\rho\\   
=\frac{1}{2} \int_{P} d \xi^+ \bar{\Theta}_-^{\rho_1 \rho} \left[ \partial_+ y_{\rho_1}     -  \partial_+ \bar{ \theta } ^\alpha ( f^{- 1 } )_{ \alpha\beta } \Psi_{\rho_1}^\beta
\right]
- \frac{1}{2} \int_{P} d \xi^- \bar{\Theta}_-^{\rho \rho_1} 
\left[    \partial_- y_{\rho_1}  +  \bar{ \Psi }_{\rho_1}^\alpha ( f^{- 1 } )_{ \alpha\beta } \partial_- \theta^\beta
\right] .
\end{gathered}
\end{align}

Using (\ref{eq of motion for v_+}) and (\ref{eq of motion for v_-}) and inserting them into (\ref{Auxiliary action S_{aux}}), we get T-dual action

\begin{align}    \label{S_{T-dual}}
\begin{gathered}
S_{T-dual}    =     k \int_{P} d^2 \xi  \Bigg[
\frac{1}{4} \bar{\Theta}_-^{\mu \nu}  \partial_+ y_\mu    \partial_- y_\nu \Bigg.  \\
\begin{aligned}
&+\frac{1}{8}  \bar{\Theta}_-^{\mu \mu_1} \bar{ \Psi }_{\mu_1}^\alpha   ( f^{- 1 } )_{ \alpha\alpha_1 }   C_\rho^{\alpha_1 \beta_1} \Delta V^\rho ( f^{- 1 } )_{ \beta_1\beta }     \Psi_{\nu_1}^{\beta} \bar{\Theta}_-^{\nu_1 \nu}  \partial_+ y_\mu    \partial_- y_\nu  \\
&  
+ \frac{1}{2} \partial_+ \bar{ \theta }^\alpha \Big( \Big. (F^{-1} ( \Delta V  ) )_{\alpha\beta}    + \frac{1}{2} ( f^{- 1 } )_{ \alpha\alpha_1 }   C_\rho^{\alpha_1 \alpha_2} \Delta V^\rho ( f^{- 1 } )_{ \alpha_2\alpha_3 }   \Psi_\mu^{ \alpha_3 }  \bar{\Theta}_-^{\mu \nu} \bar{ \Psi }_{\nu}^{\beta_1} ( f^{- 1 } )_{\beta_1\beta}   \\
&\quad - \frac{1}{2} ( f^{- 1 } )_{\alpha\alpha_1}  \Psi_\mu^{ \alpha_1 }  \bar{\Theta}_-^{\mu \nu} \bar{ \Psi }_{\nu}^{\beta_1} ( f^{- 1 } )_{\beta_1\beta} 
+ \frac{1}{2} ( f^{- 1 } )_{\alpha\alpha_1}  \Psi_\mu^{ \alpha_1 }  \bar{\Theta}_-^{\mu \nu} \bar{ \Psi }_{\nu}^{\beta_3} ( f^{- 1 } )_{ \beta_3\beta_2 }   C_\rho^{\beta_2 \beta_1} \Delta V^\rho ( f^{- 1 } )_{ \beta_1\beta }  \\
&\quad\Big. -\frac{1}{4}    ( f^{- 1 } )_{ \alpha\alpha_1 } \Psi_\mu^{ \alpha_1 } \bar{\Theta}_-^{\mu \mu_1} \bar{ \Psi }_{\mu_1}^{\alpha_2}
( f^{- 1 } )_{ \alpha_2\alpha_3 }  C_\rho^{\alpha_3 \beta_3}
\Delta V^\rho
( f^{- 1 } )_{ \beta_3\beta_2 }  \Psi_{\nu_1}^{ \beta_2 } \bar{\Theta}_-^{\nu_1  \nu} \bar{ \Psi }_\nu^{\beta_1}  ( f^{- 1 } )_{ \beta_1\beta }
\Big) \partial_- \theta^\beta    \\
& +\frac{1}{4}  \partial_+ y_\mu   \bar{\Theta}_-^{\mu  \mu_1}  \bar{ \Psi }_{\mu_1}^\alpha  \Big( (F^{-1} ( \Delta V  ) )_{\alpha\beta}    +   \frac{1}{2}   ( f^{- 1 } )_{ \alpha\alpha_1 }   C_\rho^{\alpha_1 \beta_3} 
\Delta V^\rho ( f^{- 1 } )_{ \beta_3\beta_2 } \Psi_{\nu_1}^{\beta_2}   \bar{\Theta}_-^{\nu_1  \nu} \bar{ \Psi }_\nu^{\beta_1}  ( f^{- 1 } )_{ \beta_1\beta } 
\Big)   \\
&\quad\times \partial_- \theta^\beta  \\
&- \frac{1}{4} \partial_+ \bar{ \theta }^\alpha        \Big(  (F^{-1} ( \Delta V  ) )_{\alpha\beta}  +\frac{1}{2}  ( f^{- 1 } )_{ \alpha\alpha_1 } \Psi_\mu^{\alpha_1}  \bar{\Theta}_-^{\mu \mu_1} \bar{ \Psi }_{\mu_1}^{\alpha_2}
( f^{- 1 } )_{ \alpha_2\alpha_3 } C_\rho^{\alpha_3 \beta_1} \Delta V^\rho ( f^{- 1 } )_{ \beta_1\beta } \Big)  \Psi_{\nu_1}^\beta \bar{\Theta}_-^{\nu_1 \nu}   \\
&\quad\Bigg. \times   \partial_- y_\nu  \Bigg].
\end{aligned} 
\end{gathered}
\end{align}
Let us note that above we kept terms up to to the first order in $C^{\alpha\beta}_\mu$.

T-dual action contains all terms as initial action (\ref{Action final S}) up to the change $x^\mu\to y_\mu$. Consequently, T-dual background fields are of the form

\begin{align}\label{Background fields transformations}
\begin{gathered}
{}^\star\Pi_{ +}^{\mu \nu } = \frac{1}{4} \bar{\Theta}_-^{\mu \nu} + \frac{1}{8} \bar{\Theta}_-^{\mu \mu_1} \bar{ \Psi }_{\mu_1}^\alpha \Big[\Big. (F^{-1} ( \Delta V  ) )_{\alpha\beta} +  ( f^{- 1 } )_{ \alpha\alpha_1 }   C_\rho^{\alpha_1 \beta_1} \Delta V^\rho ( f^{- 1 } )_{ \beta_1\beta } \\
\begin{aligned}
&- \frac{1}{2} ( f^{- 1 } )_{\alpha\alpha_1}  \Psi_{\mu_2}^{ \alpha_1 }  \bar{\Theta}_-^{\mu_2 \nu_2} \bar{ \Psi }_{\nu_2}^{\beta_1} ( f^{- 1 } )_{\beta_1\beta} + \frac{1}{2} ( f^{- 1 } )_{ \alpha\alpha_1 }   C_\rho^{\alpha_1 \alpha_2} \Delta V^\rho ( f^{- 1 } )_{ \alpha_2\alpha_3 }   \Psi_{\mu_2}^{ \alpha_3 }  \bar{\Theta}_-^{\mu_2 \nu_2} \bar{ \Psi }_{\nu_2}^{\beta_1} ( f^{- 1 } )_{\beta_1\beta} \\
&+\frac{1}{2} ( f^{- 1 } )_{\alpha\alpha_1}  \Psi_{\mu_2}^{ \alpha_1 }  \bar{\Theta}_-^{\mu_2 \nu_2} \bar{ \Psi }_{\nu_2}^{\beta_3} ( f^{- 1 } )_{ \beta_3\beta_2 }   C_\rho^{\beta_2 \beta_1} \Delta V^\rho ( f^{- 1 } )_{ \beta_1\beta } \\
&-\frac{1}{4}    ( f^{- 1 } )_{ \alpha\alpha_1 } \Psi_{\mu_2}^{ \alpha_1 } \bar{\Theta}_-^{\mu_2 \mu_3} \bar{ \Psi }_{\mu_3}^{\alpha_2}
( f^{- 1 } )_{ \alpha_2\alpha_3 }  C_\rho^{\alpha_3 \beta_3}
\Delta V^\rho
( f^{- 1 } )_{ \beta_3\beta_2 }  \Psi_{\nu_3}^{ \beta_2 } \bar{\Theta}_-^{\nu_3  \nu_2} \bar{ \Psi }_{\nu_2}^{\beta_1}  ( f^{- 1 } )_{ \beta_1\beta } \Big.\Big] \Psi_{\nu_1}^\beta \bar{\Theta}_-^{\nu_1 \nu},
\end{aligned}
\end{gathered}\\
\label{Background fields transformation 2}
\begin{gathered}
{}^\star(F^{-1} ( x  ) )_{\alpha\beta}  = (F^{-1} ( \Delta \bar{ y }  ) )_{\alpha\beta}  + \frac{1}{2} ( f^{- 1 } )_{ \alpha\alpha_1 }   C_\rho^{\alpha_1 \alpha_2} \Delta V^\rho ( f^{- 1 } )_{ \alpha_2\alpha_3 }   \Psi_\mu^{ \alpha_3 }  \bar{\Theta}_-^{\mu \nu} \bar{ \Psi }_{\nu}^{\beta_1} ( f^{- 1 } )_{\beta_1\beta} \\
\begin{aligned}
&- \frac{1}{2} ( f^{- 1 } )_{\alpha\alpha_1}  \Psi_\mu^{ \alpha_1 }  \bar{\Theta}_-^{\mu \nu} \bar{ \Psi }_{\nu}^{\beta_1} ( f^{- 1 } )_{\beta_1\beta} + \frac{1}{2} ( f^{- 1 } )_{\alpha\alpha_1}  \Psi_\mu^{ \alpha_1 }  \bar{\Theta}_-^{\mu \nu} \bar{ \Psi }_{\nu}^{\beta_3} ( f^{- 1 } )_{ \beta_3\beta_2 }   C_\rho^{\beta_2 \beta_1} \Delta V^\rho ( f^{- 1 } )_{ \beta_1\beta } \quad\ \  \\
&-\frac{1}{4}    ( f^{- 1 } )_{ \alpha\alpha_1 } \Psi_\mu^{ \alpha_1 } \bar{\Theta}_-^{\mu \mu_1} \bar{ \Psi }_{\mu_1}^{\alpha_2}
( f^{- 1 } )_{ \alpha_2\alpha_3 }  C_\rho^{\alpha_3 \beta_3}
\Delta V^\rho
( f^{- 1 } )_{ \beta_3\beta_2 }  \Psi_{\nu_1}^{ \beta_2 } \bar{\Theta}_-^{\nu_1  \nu} \bar{ \Psi }_\nu^{\beta_1}  ( f^{- 1 } )_{ \beta_1\beta },
\end{aligned}
\end{gathered}
\end{align} 
\begin{equation}
{}^\star \bar{ \Psi }^{\mu\alpha} = \frac{1}{2} \Theta_-^{\mu\nu}\bar{ \Psi }_\nu^\alpha, \qquad {}^\star \Psi_{\nu\beta} = -\frac{1}{2}\Psi_\mu^\beta \Theta_-^{\mu\nu}.
\end{equation}  
Comparing background field of T-dual theory with background fields from \cite{nasrad3} we immediately notice that background fields have become more complex. However, this is just an illusion. In both cases background field are exactly the same only difference is that here we did not introduce tensor $\breve{\Pi}_{+\mu\nu} = \Pi_{+\mu\nu} + \frac{1}{2}\bar{ \Psi }^\alpha_\mu F^{-1}(\Delta V) _{\alpha \beta} \Psi^\beta_\nu$    and its inverse, therefore we are missing ingredients to express our fields in more compactified format.

\subsection{T-dualization of T-dual theory}

Since the initial theory is not symmetric under translations, T-dual action that is obtained from auxiliary action (\ref{Auxiliary action S_{aux}}) and it is now invariant to translations of T-dual coordinates. Consequently, we can dualize T-dual theory by generalized Buscher procedure. We start with the introduction of following substitutions

\begin{align}
\partial_\pm y_\mu       \rightarrow       &   D_\pm y_\mu  = \partial_\pm y_\mu +  u_{\pm \mu}  \rightarrow D_\pm y_\mu =  u_{\pm \mu}, \\
\Delta \bar{ y }^\rho       \rightarrow     &   \Delta \bar{ u }^\rho, \\
\Delta \bar{ u }^\rho  = 
&\frac{1}{2} \int_{P} d \xi^+ \bar{\Theta}_-^{\rho_1 \rho} \left[ u_{+ \rho_1}     -  \partial_+ \bar{ \theta } ^\alpha ( f^{- 1 } )_{ \alpha\beta } \Psi_{\rho_1}^\beta
\right] \nonumber \\
- &\frac{1}{2} \int_{P} d \xi^- \bar{\Theta}_-^{\rho \rho_1} 
\left[    u_{- \rho_1}  +  \bar{ \Psi }_{\rho_1}^\alpha ( f^{- 1 } )_{ \alpha\beta } \partial_- \theta^\beta 
\right]\, ,\\ 
S \rightarrow & S + \frac{1}{2}   (u_{+ \mu}  \partial_-x^\mu  - u_{- \mu}\partial_+ x^\mu  ) \, .
\end{align}
From the first line we see that gauge is fixed choosing $y(\xi) = const$. Inserting these substitutions into (\ref{S_{T-dual}}) we obtain

\begin{align}    \label{S_{gauge fix}}
\begin{gathered}
S_{gauge fix}    =     \kappa \int_{P} d^2 \xi  \Bigg[
\frac{1}{4} \bar{\Theta}_-^{\mu \nu}  u_{ + \mu}    u_{- \nu} \Bigg.  \\
\begin{aligned}
&+\frac{1}{8}  \bar{\Theta}_-^{\mu \mu_1} \bar{ \Psi }_{\mu_1}^\alpha   ( f^{- 1 } )_{ \alpha\alpha_1 }   C_\rho^{\alpha_1 \beta_1} \Delta \bar{ u }^\rho ( f^{- 1 } )_{ \beta_1\beta }     \Psi_{\nu_1}^{\beta} \bar{\Theta}_-^{\nu_1 \nu}  u_{ + \mu}    u_{- \nu}  \\
&  
+ \frac{1}{2} \partial_+ \bar{ \theta }^\alpha \Big( \Big. (F^{-1} ( \Delta \bar{ u }  ) )_{\alpha\beta}    + \frac{1}{2} ( f^{- 1 } )_{ \alpha\alpha_1 }   C_\rho^{\alpha_1 \alpha_2} \Delta \bar{ u }^\rho ( f^{- 1 } )_{ \alpha_2\alpha_3 }   \Psi_\mu^{ \alpha_3 }  \bar{\Theta}_-^{\mu \nu} \bar{ \Psi }_{\nu}^{\beta_1} ( f^{- 1 } )_{\beta_1\beta}   \\
&\quad - \frac{1}{2} ( f^{- 1 } )_{\alpha\alpha_1}  \Psi_\mu^{ \alpha_1 }  \bar{\Theta}_-^{\mu \nu} \bar{ \Psi }_{\nu}^{\beta_1} ( f^{- 1 } )_{\beta_1\beta} 
+ \frac{1}{2} ( f^{- 1 } )_{\alpha\alpha_1}  \Psi_\mu^{ \alpha_1 }  \bar{\Theta}_-^{\mu \nu} \bar{ \Psi }_{\nu}^{\beta_3} ( f^{- 1 } )_{ \beta_3\beta_2 }   C_\rho^{\beta_2 \beta_1} \Delta \bar{ u }^\rho ( f^{- 1 } )_{ \beta_1\beta }  \\
&\quad\Big. -\frac{1}{4}    ( f^{- 1 } )_{ \alpha\alpha_1 } \Psi_\mu^{ \alpha_1 } \bar{\Theta}_-^{\mu \mu_1} \bar{ \Psi }_{\mu_1}^{\alpha_2}
( f^{- 1 } )_{ \alpha_2\alpha_3 }  C_\rho^{\alpha_3 \beta_3}
\Delta \bar{ u }^\rho
( f^{- 1 } )_{ \beta_3\beta_2 }  \Psi_{\nu_1}^{ \beta_2 } \bar{\Theta}_-^{\nu_1  \nu} \bar{ \Psi }_\nu^{\beta_1}  ( f^{- 1 } )_{ \beta_1\beta }
\Big) \partial_- \theta^\beta    \\
& +\frac{1}{4}  u_{ + \mu}   \bar{\Theta}_-^{\mu  \mu_1}  \bar{ \Psi }_{\mu_1}^\alpha  \Big( (F^{-1} ( \Delta \bar{ u }  ) )_{\alpha\beta}    +   \frac{1}{2}   ( f^{- 1 } )_{ \alpha\alpha_1 }   C_\rho^{\alpha_1 \beta_3} 
\Delta \bar{ u }^\rho ( f^{- 1 } )_{ \beta_3\beta_2 } \Psi_{\nu_1}^{\beta_2}   \bar{\Theta}_-^{\nu_1  \nu} \bar{ \Psi }_\nu^{\beta_1}  ( f^{- 1 } )_{ \beta_1\beta } 
\Big)   \\
&\quad\times \partial_- \theta^\beta  \\
&- \frac{1}{4} \partial_+ \bar{ \theta }^\alpha        \Big(  (F^{-1} ( \Delta \bar{ u }  ) )_{\alpha\beta}  +\frac{1}{2}  ( f^{- 1 } )_{ \alpha\alpha_1 } \Psi_\mu^{\alpha_1}  \bar{\Theta}_-^{\mu \mu_1} \bar{ \Psi }_{\mu_1}^{\alpha_2}
( f^{- 1 } )_{ \alpha_2\alpha_3 } C_\rho^{\alpha_3 \beta_1} \Delta \bar{ u }^\rho ( f^{- 1 } )_{ \beta_1\beta } \Big)  \Psi_{\nu_1}^\beta \bar{\Theta}_-^{\nu_1 \nu}   \\
&\quad \times   u_{- \nu}\\
 &+ \frac{1}{2}   (u_{+ \mu}  \partial_-x^\mu  - u_{- \mu}\partial_+ x^\mu  )  \Bigg. \Bigg].
\end{aligned} 
\end{gathered}
\end{align}
Using equations of motion for Lagrange multipliers, we return to the T-dual action. Finding equations of motion for gauge fields, we have

\begin{align}  
\label{T^2 equations of motion 1}
u_{+ \mu}(\kappa)   =  &\phantom{-\ }2\bar{\Pi}_{+ \nu \mu} \partial_+ x^\nu (\kappa)
-  \partial_+ x^\nu (\kappa) \bar{ \Psi }_\nu^\alpha  (f^{- 1 } )_{ \alpha\alpha_1 }     C_\rho^{\alpha_1 \beta_1}  \Delta x^\rho (f^{- 1 } )_{ \beta_1\beta } \Psi_\mu^\beta      \nonumber \\  
&+ \partial_+ \bar{ \theta }^\alpha (\kappa) (F^{-1} ( \Delta \bar{ x }  ))_{\alpha \beta} \Psi_\mu^\beta \nonumber  \\
&- \int_{\Sigma} d^2 \xi \Big(
\partial_+ \bar{ \theta }^\alpha (\xi) +\partial_+x^{\mu_1} (\xi) \bar{ \Psi }_{\mu_1}^\alpha
\Big) 
( f^{ -1 } )_{ \alpha\alpha_1 }     C_\mu^{\alpha_1 \beta_1}      ( f^{ -1 } )_{ \beta_1\beta } N(\kappa^-) \nonumber \\
&\quad\times
\Big(
\partial_- \theta^\beta (\xi) + \Psi_{\nu}^\beta \partial_- x^\nu (\xi)
\Big), \\
\label{T^2 equations of motion 2}
u_{- \nu}(\kappa)    =  &-2 \bar{\Pi}_{+\nu \mu}\partial_- x^\mu (\kappa)
+ \bar{ \Psi }_\nu^\alpha (f^{- 1 } )_{ \alpha\alpha_1 }     C_\rho^{\alpha_1 \beta_1}  \Delta x^\rho (f^{- 1 } )_{ \beta_1\beta } \Psi_\mu^\beta \partial_- x^\mu (\kappa)  \nonumber\\
&- \bar{ \Psi }_\nu^\alpha (F^{-1} ( \Delta \bar{ x }  ))_{\alpha \beta} \partial_- \theta^\beta  (\kappa) \nonumber \\
&+ \int_{\Sigma} d^2 \xi     
\Big(
\partial_+ \bar{ \theta }^\alpha (\xi) +\partial_+x^{\mu} (\xi) \bar{ \Psi }_{\mu}^\alpha
\Big) 
( f^{ -1 } )_{ \alpha\alpha_1 }     C_\nu^{\alpha_1 \beta_1}      ( f^{ -1 } )_{ \beta_1\beta } N(\kappa^-) \nonumber \\
&\quad\times
\Big(
\partial_- \theta^\beta (\xi) + \Psi_{\nu_1}^\beta \partial_- x^{\nu_1} (\xi)
\Big)\, .
\end{align}
Here we have that $\Delta x^\mu = x(\xi) - x(\xi_0)$, and inserting these equations into the gauge fixed action, keeping all terms linear with respect to $C_\rho^{\mu\nu}$ and selecting $\xi_0$ such that $x(\xi_0) = 0$, we obtain our original action (\ref{Action final S}).

\section{Non-commutative relations}
\setcounter{equation}{0}

In this chapter we will establish a relationship between Poisson brackets of original and T-dual theory using results from the previous one. Original theory is a geometric one, which means that canonical variables $x^\mu(\xi)$ and $\pi_\mu(\xi)$ satisfy standard Poisson algebra

\begin{equation}
\{ x^\mu (\sigma), \pi_\nu (\bar{\sigma})  \} = \delta^\mu_\nu \delta(\sigma - \bar{\sigma}), \quad \{ x^\mu(\sigma) , x^\nu(\bar{\sigma}) \} = 0, \quad \{ \pi_\mu(\sigma), \pi_\nu(\bar{\sigma})  \} = 0.
\end{equation}
We will find Poisson structure of T-dual theory using relations (\ref{eq of motion for v_+}) and (\ref{eq of motion for v_-}) and expressing them in terms of the coordinates and momenta of the initial theory. Replacing gauge fields with solutions of equations of motion for Lagrange multipliers, we get T-dual transformation laws in Lagrangian form. Because we implement here canonical approach, the next step is removing of all terms that are proportional to $\partial_\tau x^\mu(\xi)$. The most of the terms of this type will get incorporated into expression for canonical momenta $\pi_\mu(\xi)$, but term that is non-local and which is dependent on function $N(\xi^\pm)$ remains. One way of removing this term is to first use equations of motion for coordinate $x^\mu(\xi)$, and then replace remaining $\partial_\tau x^\mu$ term with canonical momentum.  By doing all the steps that were outlined, we have following relationship between T-dual coordinate and variables of starting theory

\begin{align} \label{sigma derivative of y}
\begin{gathered}
\partial_{\sigma} y_\nu (\sigma) =2B_{\nu \mu} \partial_{\sigma} x^\mu - G_{\nu \mu} 
(\bar{\Pi}_+ +\bar{\Pi}_+^T)^{-1 \nu_1 \mu}
\Bigg[
\frac{\pi_{\nu_1}}{k}    - \frac{1}{2} \bar{ \Psi }_{\nu_1}^\alpha \left( F^{-1} (x) \right)_{\alpha \beta} \partial_- \theta^\beta  \Bigg. \\
\begin{aligned}
& - \frac{1}{2} \partial_+ \bar{ \theta }^\alpha \left( F^{-1} (x) \right)_{\alpha \beta} \Psi_{\nu_1}^\beta    - \Big[ \Pi_{+ \mu_1 \mu_2} + \frac{1}{2} \bar{ \Psi }_{\mu_1}^\alpha \left( F^{-1} (x) \right)_{\alpha \beta} \Psi_{\mu_2}^\beta  
\Big]       (\delta_{\nu_2}^{\mu_1} \delta_{\nu_1}^{\mu_2} - \delta_{\nu_1}^{\mu_1} \delta_{\nu_2}^{\mu_2}   )  \partial_{\sigma}x^{\nu_2}  \\
&+\frac{1}{2}
\bar{ \Psi }_{\mu_1}^\alpha ( f^{ -1 } )_{ \alpha\alpha_1 }     C_\rho^{\alpha_1 \beta_1}   x^\rho(\sigma)   ( f^{ -1 } )_{ \beta_1\beta }   \Psi_{\mu_2}^\beta (\delta_{\nu_2}^{\mu_1} \delta_{\nu_1}^{\mu_2} - \delta_{\nu_1}^{\mu_1} \delta_{\nu_2}^{\mu_2}   )    (\bar{\Pi}_+ +\bar{\Pi}_+^T)^{-1 \rho \nu_2}   \\
&\quad\times
\Big[ \frac{\pi_\rho}{k} - \frac{1}{2} \bar{ \Psi }_\rho^\gamma (f^{-1})_{\gamma \gamma_1} \partial_- \theta^{\gamma_1} - \frac{1}{2} \partial_+ \bar{ \theta }^\gamma (f^{-1})_{\gamma \gamma_1} \Psi_\rho^{\gamma_1} + \bar{\Pi}_{+\rho \rho_1} \partial_{\sigma} x^{\rho_1 } - \bar{\Pi}_{+\rho_1 \rho} \partial_{\sigma} x^{\rho_1 } 
\Big]
\Bigg].
\end{aligned}
\end{gathered}
\end{align}

To find Poisson bracket between T-dual coordinates, we can start by finding Poisson bracket of sigma derivatives of T-dual coordinates and then integrating twice (see \cite{nasrad3}). Implementing this procedure we have that Poisson bracket for sigma derivatives is given as

\begin{align} \label{sigma derivative Poisson bracket}
\begin{gathered}
\{\partial_{\sigma_1} y_{\nu_1} (\sigma_1), \partial_{\sigma_2} y_{\nu_2} (\sigma_2)  \} = \\
\begin{aligned}
&= \frac{2}{k} (\bar{\Pi}_+ + \bar{\Pi}_+^T)^{-1 \mu_1 \mu_2} \Big[ G_{\nu_1 \mu_1} B_{\nu_2 \mu_2} \partial_{\sigma_2} \delta(\sigma_1 - \sigma_2)    -   B_{\nu_1 \mu_1}G_{\nu_2\mu_2} \partial_{\sigma_1}     \delta(\sigma_1- \sigma_2)
\Big]  \\
&+ \frac{1}{k} \bar{ \Psi }_{\nu_3}^\alpha (f^{-1})_{\alpha \alpha_1} C_\rho^{ \alpha_1 \beta_1} (f^{-1})_{\beta_1 \beta} \Psi_{\nu_4}^\beta ( \delta_{\mu_3}^{\nu_3} \delta_{\mu_4}^{\nu_4}  + \delta_{\mu_3}^{\nu_4}\delta_{\mu_4}^{\nu_3}     ) (\bar{\Pi}_+ + \bar{\Pi}_+^T)^{-1 \mu_3 \mu_1} (\bar{\Pi}_+ + \bar{\Pi}_+^T)^{-1 \mu_4 \mu_2}  \\
&\quad\times \Big[  
G_{\nu_1 \mu_1} B_{\nu_2 \mu_2}  x^\rho (\sigma_1) \partial_{\sigma_2} \delta(\sigma_1 - \sigma_2)    -   B_{\nu_1 \mu_1}G_{\nu_2\mu_2} x^\rho (\sigma_2) \partial_{\sigma_1}     \delta(\sigma_1- \sigma_2) 
\Big].
\end{aligned}
\end{gathered}
\end{align}
Then we integrate with respect to $\sigma_1$ ($\sigma_2$), where we set boundaries as $\sigma_0$ ($\bar{\sigma}_0$) and $\sigma$ ($\bar{\sigma}$). Extracting only Poisson bracket terms that contain $\sigma $ and $\bar{\sigma}$, we have

\begin{align} \label{Poisson brackets T-dual}
\begin{gathered}
\{ y_{\nu_1} (\sigma),  y_{\nu_2} (\bar{\sigma})  \} = \frac{2}{k} (\bar{\Pi}_+ + \bar{\Pi}_+^T)^{-1 \mu_1 \mu_2} \Big[ G_{\nu_1 \mu_1} B_{\nu_2 \mu_2}     +   B_{\nu_1 \mu_1}G_{\nu_2\mu_2} 
\Big] H(\sigma - \bar{\sigma}) \\
\begin{aligned}
&+ \frac{1}{k}\bar{ \Psi }_{\nu_3}^\alpha (f^{-1})_{\alpha \alpha_1} G_\rho^{ \alpha_1 \beta_1} (f^{-1})_{\beta_1 \beta} \Psi_{\nu_4}^\beta ( \delta_{\mu_3}^{\nu_3} \delta_{\mu_4}^{\nu_4}  + \delta_{\mu_3}^{\nu_4}\delta_{\mu_4}^{\nu_3}     ) (\bar{\Pi}_+ + \bar{\Pi}_+^T)^{-1 \mu_3 \mu_1} (\bar{\Pi}_+ + \bar{\Pi}_+^T)^{-1 \mu_4 \mu_2} \\
&\times    \Big[ G_{\nu_1 \mu_1} B_{\nu_2 \mu_2} x^\rho (\bar{\sigma})    +   B_{\nu_1 \mu_1}G_{\nu_2\mu_2} x^\rho (\sigma)
\Big]H(\sigma - \bar{\sigma}) .
\end{aligned}
\end{gathered}
\end{align}
Here, $H(\sigma - \bar{\sigma})$ is same step function defined in Appendix \ref{appendix A}. It should be noted that these Poisson brackets are zero when $\sigma = \bar{\sigma}$. However, in cases where string in curled around compactified dimension, that is cases where $\sigma - \bar{\sigma} = 2\pi$, we have following situation

\begin{align}
\begin{gathered}
\{ y_{\nu_1} (\sigma + 2\pi),  y_{\nu_2} (\sigma)  \} =\frac{2}{k} (\bar{\Pi}_+ + \bar{\Pi}_+^T)^{-1 \mu_1 \mu_2} \Big[ G_{\nu_1 \mu_1} B_{\nu_2 \mu_2}     +   B_{\nu_1 \mu_1}G_{\nu_2\mu_2} 
\Big] \\
\begin{aligned}
&+ \frac{1}{k}\bar{ \Psi }_{\nu_3}^\alpha (f^{-1})_{\alpha \alpha_1} C_\rho^{ \alpha_1 \beta_1} (f^{-1})_{\beta_1 \beta} \Psi_{\nu_4}^\beta ( \delta_{\mu_3}^{\nu_3} \delta_{\mu_4}^{\nu_4}  + \delta_{\mu_3}^{\nu_4}\delta_{\mu_4}^{\nu_3}     ) (\bar{\Pi}_+ + \bar{\Pi}_+^T)^{-1 \mu_3 \mu_1} (\bar{\Pi}_+ + \bar{\Pi}_+^T)^{-1 \mu_4 \mu_2} \\
&\times    \Big[4\pi  G_{\nu_1 \mu_1} B_{\nu_2 \mu_2}  N^\rho +   \big(  G_{\nu_1 \mu_1} B_{\nu_2 \mu_2}    +   B_{\nu_1 \mu_1}G_{\nu_2\mu_2}  \big) x^\rho (\sigma)
\Big]    .
\end{aligned}
\end{gathered}
\end{align}
We used fact that $H(2\pi) = 1$. The symbol $N^\mu$ denotes winding number around compactified coordinate, if is defined as

\begin{equation}
x^\mu(\sigma + 2 \pi) - x^\mu(\sigma) = 2\pi N^\mu.
\end{equation}
Let us note that if we choose $x^\mu(\sigma) = 0$ than Poisson bracket has linear dependence on winding number. In cases where we don't have any winding number, we still have non-commutativity that is proportional to background fields.

Using the expression for sigma derivative of $y_\nu$ (\ref{sigma derivative of y}) and expression for Poisson bracket of sigma derivatives (\ref{sigma derivative Poisson bracket}), we can find non-associative relations. Procedure is the same as for finding Poisson brackets of T-dual theory, we find Poisson bracket of sigma derivatives and integrate with respect to sigma coordinate, this time integral is done trice. Going along with this procedure we have following final result

\begin{align}
\begin{gathered}
\{  y_\nu (\sigma)  ,  \{ y_{\nu_1} (\sigma_1),  y_{\nu_2} (\sigma_2)  \} \} = \frac{G_{\nu \mu}}{k^2} (\bar{\Pi}_+ + \bar{\Pi}_+^T)^{-1 \rho \mu} \\
\begin{aligned}
&\times \bar{ \Psi }_{\nu_3}^\alpha (f^{-1})_{\alpha \alpha_1} C_\rho^{ \alpha_1 \beta_1} (f^{-1})_{\beta_1 \beta} \Psi_{\nu_4}^\beta( \delta_{\mu_3}^{\nu_3} \delta_{\mu_4}^{\nu_4}  + \delta_{\mu_3}^{\nu_4}\delta_{\mu_4}^{\nu_3}     ) (\bar{\Pi}_+ + \bar{\Pi}_+^T)^{-1 \mu_3 \mu_1}   (\bar{\Pi}_+ + \bar{\Pi}_+^T)^{-1 \mu_4 \mu_2}
\\
& \times\Big[ G_{\nu_1 \mu_1} B_{\nu_2 \mu_2} H(\sigma - \sigma_2)    +   B_{\nu_1 \mu_1}G_{\nu_2\mu_2} H(\sigma - \sigma_1)
\Big]H(\sigma_1 - \sigma_2) .
\end{aligned}
\end{gathered}
\end{align}  
Since Jacobi identity is non-zero for T-dual theory we have that coordinate dependent RR field produces non-associative theory. However putting $\sigma_1=\sigma_2=\bar{\sigma}$ and $\sigma= \bar{\sigma} +2\pi$ we have that Jacobi identity disappears

\begin{equation}
\{  y_\nu (\bar{\sigma} + 2\pi)  ,  \{ y_{\nu_1} (\bar{\sigma}),  y_{\nu_2} (\bar{\sigma})  \} \} = 0.
\end{equation}

\section{Conclusion}
\setcounter{equation}{0}

In this article we examined type II superstring propagating in presence of coordinate dependent RR field. This choice of background was in accordance with consistency conditions for background field and all calculations were made in approximation that are linear with respect to coordinate dependent part of RR field. We have also excluded parts that were non-linear in fermionic coordinates and neglected pure spinor actions. Using equations of motion for fermionic momenta we obtained action that was expressed in terms of bosonic coordinates, their derivatives and derivatives of fermionic coordinates. Unlike \cite{nasrad3} we do not impose any conditions on the constant and coordinate dependent part of RR field strength, so, it is not possible to deduce this case to the form of the weakly curved background one.

Action with our choice of background fields did not possess translation symmetry, therefore we needed to use Buscher procedure that was extended to such cases. By substituting starting action with auxiliary action we gave up on locality in order to be able to find T-dual theory. Finding equations of motion of newly introduced Lagrange multipliers we were able to salvage starting action giving us assurance that auxiliary action we selected would produce correct T-dual theory. After this we found equations of motion for gauge fields and by inserting them into action, we found T-dual theory. 

Having found T-dual theory, we applied T-dual procedure once again as a more thorough way of checking if action we obtained was in fact correct T-dual of starting action. Unlike starting action, T-dual action possessed translation symmetry and was non-local from the start by virtue of having dual coordinate $\Delta V^\mu$. Applying steps of generalized Buscher procedure \cite{Generalized Buscher procedure, Auxiliary action Buscher procedure} we obtained starting action, again confirming that our choice of auxiliary action was correct.

We obtained non-commutativity relations in context of T-dual theory, where we used T-dual transformation laws as a bridge between Poisson brackets of starting theory and T-dual theory. T-dual transformation laws were expressed as functions of coordinates and momenta of original theory and using their standard Poisson algebra, we got non-commutativity in T-dual theory. From expression for Poisson brackets (\ref{Poisson brackets T-dual}) we can see that non-commutativity is proportional to infinitesimal part of RR field as well as to symmetrised inverse of field $\bar{\Pi}$. Non-commutativity relations are zero in case when $\sigma = \bar{\sigma}$, while in case where $\sigma = \bar{\sigma} + 2\pi$ we see the emergence of winding numbers.

Taking into account Poisson brackets of sigma derivatives and expression for sigma derivative of T-dual coordinate we were able to find non-associative relation for T-dual theory. In general case this relation was non-zero and it was proportional to infinitesimal constant $C_\rho^{\mu \nu}$. In special case when we put  $\sigma_1=\sigma_2=\bar{\sigma}$ and $\sigma= \bar{\sigma} +2\pi$ we noticed that non-associativity relation disappears. During the implementation of the T-dualization procedure and calculations, we obtained generalization of $\beta_\mu$ functions in the form of the $N$-functions.

It should be noted that since we did not preform T-dualization along fermionic coordinates their Poisson structure would remain the same as in original theory. Furthermore, since background fields do not depend on fermionic coordinates it should be expected, as in the case of bosonic coordinates \cite{Generalized Buscher procedure 3}, that T-duality would leave Poisson brackets between fermionic fields the same. We expect that, if proposed non-commutative relations from \cite{Vertex operators, original paper 2} are even possible, we would need at least RR field that depends both on fermionic and bosonic coordinates.

\appendix
\setcounter{equation}{0}
\section{Obtaining $N(\kappa^\pm)$ terms}
\label{appendix A}

In this paper function $N(\kappa^\pm)$ emerged in T-dual transformation laws as a consequence of variation of term that was proportional to $\Delta V$. Here we will present derivation of this function. 

\begin{align}
\begin{gathered}
\frac{ \delta \left( F^{-1}  ( \Delta V )   \right)_{ \alpha\beta } }{ \delta v_+^\mu ( \kappa)} =
- ( f^{ -1 } )_{ \alpha\alpha_1 }     C_l^{\alpha_1 \beta_1}      ( f^{ -1 } )_{ \beta_1\beta }  
\int_{P} d { \xi^\prime }^m  \frac{\delta v_m^\rho ( \xi^\prime )}{ \delta v_+^\mu (\kappa) } = \\
\begin{aligned}
&=- ( f^{ -1 } )_{ \alpha\alpha_1 }     C_\mu^{\alpha_1 \beta_1}      ( f^{ -1 } )_{ \beta_1\beta }  
\int_{P} d { \xi^\prime }^+ \delta( { \xi^\prime }^+   - \kappa^+ )       \delta(   { \xi^\prime }^-   - \kappa^-  ) \\
& = - ( f^{ -1 } )_{ \alpha\alpha_1 }     C_\mu^{\alpha_1 \beta_1}      ( f^{ -1 } )_{ \beta_1\beta }  
\int_{t_i}^{t_f} dt \frac{d{ \xi^\prime }^+}{dt} \delta( { \xi^\prime (t) }^+   - \kappa^+ )       \delta(   { \xi^\prime }^- (t)   - \kappa^-  ) \\
& = - ( f^{ -1 } )_{ \alpha\alpha_1 }     C_\mu^{\alpha_1 \beta_1}      ( f^{ -1 } )_{ \beta_1\beta }  
\int_{{\xi_0 }^+}^{{\xi }^+} du  \delta( u   - \kappa^+ )       \delta      \Big(      { \xi^\prime }^- \big( ( { { \xi^\prime }^+ })^{-1}
(u) \big)   - \kappa^-
\Big)    \\
&= - ( f^{ -1 } )_{ \alpha\alpha_1 }     C_\mu^{\alpha_1 \beta_1}      ( f^{ -1 } )_{ \beta_1\beta } 
\delta      \Big(      { \xi^\prime }^- \big( ( { { \xi^\prime }^+ })^{-1}
(\kappa^+) \big)   - \kappa^-
\Big)     
\left[    H( \xi^+   - \kappa^+  )  -  H(\xi_0^+    -   \kappa^+)    \right]\\
&=- ( f^{ -1 } )_{ \alpha\alpha_1 }     G_\mu^{\alpha_1 \beta_1}      ( f^{ -1 } )_{ \beta_1\beta } N(\kappa^+).
\end{aligned}
\end{gathered}
\end{align}

In third line we have parametrized the path with parameter $t$ where ${\xi^\prime }^+(t_i) ={\xi_0 }^+$ and ${\xi^\prime }^+(t_f) ={\xi }^+$. In fourth line we introduced substitution $u={\xi^\prime }^+(t)$, in delta function this substitute is inverted. Fifth line is obtained by using following integration rule for Dirac delta function

\begin{equation}
\int_{\sigma_0}^{\sigma} d\eta f(\eta) \delta(\eta - \bar{\eta}) = f(\bar{\eta}) \left[ H(\sigma - \bar{\eta})  - H(\sigma_0 - \bar{\eta})   \right].
\end{equation}
 
Here, $H(x)$ is a step function defined as

\begin{align}
H(x) &= \int_0^x d\eta \delta(\eta) = \frac{1}{2\pi} \Big[x + \sum_{n \geq 2} \frac{1}{n} \sin (nx)  \Big] \nonumber \\
&= \begin{cases}
\text{$0$   \ \ \ \ \ \  if\ \ \  $x = 0$}\\
\text{$1/2$  \ \ \  if\ \ \  $0 < x < 2\pi$}\\
\text{$1$   \ \ \  \ \ \ if\ \ \  $x = 2\pi$}
\end{cases}.
\end{align}

Procedure for obtaining $N(\kappa^-)$ is similar.

\setcounter{equation}{0}
\section{Properties of $N(\kappa^\pm)$ terms}
\label{appendix B}

Here ve will list some properties of  $N(\kappa^\pm)$ function.

\begin{gather}
N(\kappa^+) + N(\kappa^-) = N(\kappa^0),\\
N(\kappa^+) - N(\kappa^-) = N(\kappa^1),
\end{gather}

Where $\kappa^0$ and $\kappa^1$ represent $\tau$ and $\sigma$ coordinates respectevly

\begin{gather}
\int_{\Sigma} d^2 \xi \partial_+ N(\kappa^+)  = 1, \qquad
\int_{\Sigma} d^2 \xi \partial_- N(\kappa^+)  = 0,\\
\int_{\Sigma} d^2 \xi \partial_- N(\kappa^-)  = 1, \qquad
\int_{\Sigma} d^2 \xi \partial_+ N(\kappa^-)  = 0.
\end{gather}
These relationships can be checked directly by applying partial derivatives to expressions from \ref{appendix A}.

\begin{gather}
\int_{\Sigma} d^2 \xi \partial_+ N(\kappa^+) = \int_{\Sigma} d^2 \xi \delta      \Big(      { \xi^\prime }^- \big( ( { { \xi^\prime }^+ })^{-1}
(\kappa^+) \big)   - \kappa^-
\Big) \partial_+    
\left[    H( \xi^+   - \kappa^+  )  -  H(\xi_0^+    -   \kappa^+)    \right]\nonumber\\
=\int_{\Sigma} d^2 \xi \delta      \Big(      { \xi^\prime }^- \big( ( { { \xi^\prime }^+ })^{-1}
(\kappa^+) \big)   - \kappa^-
\Big) \delta ( \xi^+   - \kappa^+  ) = \int d\xi^- \delta      \Big(      { \xi^\prime }^- \big( ( { { \xi^\prime }^+ })^{-1}
(\xi^+) \big)   - \kappa^-
\Big).
\end{gather}

In appendix \ref{appendix A} we had following parametrisation of path P : ${\xi^\prime }^+(t_i) ={\xi_0 }^+$ and ${\xi^\prime }^+(t_f) ={\xi }^+$. Applying inverse parametrisation we have  $({\xi^\prime }^+)^{-1}(\xi_0^+ )= t_i$ and $({\xi^\prime }^+)^{-1}(\xi^+) = t_f$. With these we have

\begin{gather}
\int d\xi^- \delta      \Big(      { \xi^\prime }^- \big( ( { { \xi^\prime }^+ })^{-1}
(\xi^+) \big)   - \kappa^-
\Big) = \int_{\Sigma} d\xi^-  \delta      \Big(      { \xi^\prime }^- \big(t_f \big)   - \kappa^-
\Big)\nonumber \\ 
= \int_{\Sigma} d\xi^- \delta \Big( \xi^- - \kappa^-   \Big) = 1.
\end{gather}

Same rules apply for $N(\kappa^-)$, $N(\kappa^0)$ and $N(\kappa^1)$. In cases where $F^{-1}(x)_{\alpha \beta}$ is antisymmetric we can transfer partial derivatives from $\partial_\pm V^mu$ to $N(\kappa^\pm)$ and obtain standard $\beta^\pm$ functions.

\end{document}